\documentclass[12pt]{article}

\begin{document}

\title{Duality transformations and temperature inversion symmetry }
\author{{\large F C Santos$^{\star}$ and A C Tort$%
^{\dagger}$} \\
Instituto de F\'{\i}sica - Universidade Federal do Rio de Janeiro - CP 68528
\\
Rio de Janeiro, RJ, Brasil - 21945-970.}
\date{\today}
\maketitle
\begin{abstract}
We argue that the temperature inversion symmetry present in the original Casimir setup and also in other Casimir systems for which symmetrical boundary conditions are imposed is not related to the duality transformations that in the context defined in \cite{Fukushima&Ohta2001}  are transformations relating spatial extension and temperature, and pressure and energy density. We provide an example of a Casimir system for which in principle there is no temperature inversion symmetry but nevertheless these duality transformations can be found.

\end{abstract}
\vfill
\noindent $^{\star }$ {e-mail: filadelf@i.f.ufrj.br}\\
\noindent $^\dagger$ {e-mail: tort@i.f.ufrj.br}
\clearpage
\section{Duality transformations and temperature inversion \\ symmetry}
In a recently published paper Fukushima and Ohta  \cite{Fukushima&Ohta2001} discussed the relationship between the Casimir effect \cite{Casimir48} and the Stefan-Boltzmmann $T^4$-law or Planck's radiation law. In particular they discussed the explict conversion from the Casimir energy density to the $T^4$-law. The bottom line in Fukushima and Ohta's paper is that there is a correspondence between the Casimir energy density at zero temperature and the Stefan-Boltzmann free energy density at $T=T_{eff}$, where $T_{eff}$ is an effective temperature that depends on the geometrical features of the experimental setup. Fukushima and Ohta correctly argued that the effective temperature cannot be regarded as a genuine temperature. This is so because the effective temperature does not satisfy the fundamental thermodynamical relation
\begin{equation}\label{trueT}
\frac{\partial s}{\partial u}=\frac{1}{T},
\end{equation}
where $s$ and $u$ are the entropy and energy density, respectively. In fact, there is not a true thermodynamics associated with the effective temperature. The effective temperature is introduced through duality transformations. The duality transformations in the context of Ref. \cite{Fukushima&Ohta2001} are obtained by writing the Casimir energy density and pressure as integrals that are formally similar to their thermodynamic counterparts. For the original Casimir's setup these transformations read 
\begin{equation}\label{dualCasimir}
2\ell \rightleftharpoons \frac{1}{T}\,; \hskip 2cm p\rightleftharpoons -u.
\end{equation}

These transformations are supposed to be related to a dual symmetry exhibited by the Helmholtz free energy. This dual symmetry is called temperature inversion symmetry and the conditions for its existence in the Casimir effect were well established by Ravndal and Tollefsen \cite{Ravndal&Tollefsen89}. In the case of the Casimir effect for two perfectly conducting square parallel plates of side $L$ kept at a distance $\ell$ from each other with $L\gg \ell$ \cite{Casimir48}, the temperature inversion symmetry can be expressed by \cite{Brown&Maclay69}
\begin{equation}\label{TIS1}
\left(2\pi\xi\right)^4F\left(\frac{1}{4\pi^2\xi}\right)=F
\left(\xi\right),
\end{equation}  
where $F$ is the free energy and $\xi:=T\ell/\pi$. 

Suppose, however, that $\xi$ assumes extremely large values. Then starting from the free energy density at zero temperature, i.e.: the Casimir energy density, and making use of (\ref{TIS1}) it is possible to obtain the Stefan-Boltzmann $T^4$ law, where $T$ is a true temperature, that is, it is not dependent on the geometrical features of the system and obeys definition (\ref{trueT}). There are, however, two physical ways of obtaining the Stefan-Boltzmann law. One is to take the very high temperature limit, $T\to\infty$, in which case the distance between the plates is kept constant and (\ref{TIS1})  yields
\begin{equation}
\frac{F\left(T\to\infty\right)}{\ell L^2}=-\left(2T\ell\right)^4\frac{\pi^2}{2^3\times90\,\ell^4}=-\frac{\pi^2T^4}{45}.
\end{equation}
The other way is to set the plates very far apart, that is, $\ell\to\infty$, while the temperature is kept constant. In this case from (\ref{TIS1}) again we obtain
\begin{equation}
\frac{F\left(T\right)}{\ell L^2}=-\left(2T\ell\right)^4\frac{\pi^2}{2^3\times90\,\ell^4}=-\frac{\pi^2T^4}{45}.
\end{equation}
Notice that in contrast with the first case, in the second case the Stefan-Boltzmann law is obtained for arbitrary temperature. If the duality transformations are supposed to be valid at any temperature, for arbitrary separation $\ell$, then as shown in Ref. \cite{Fukushima&Ohta2001} we can write the Casimir energy density as a Stefan-Boltzamnn energy density at an effective temperature. But this is in contradiction with the above remarks concerning the behavior of the free energy density under temperature inversion symmetry. For arbitrary $T$, only in the limit $\ell\to\infty$ does the free energy becomes proportional to $T^4$. The solution to this puzzle is to realize that temperature inversion symmetry and the transformations given by (\ref{dualCasimir}) are unrelated.
 
Nevertheless, the method of obtaining the Casimir energy and pressure employed by Fukushima and Ohta is interesting and valuable in itself and it is worthwhile to provide one more example of its usefulness. In the next section we will show how to obtain relations similar to (\ref{dualCasimir}) in the case of two parallel surfaces on which we impose  Dirichlet boundary conditions on one surface and Neumann boundary conditions on the other. Since the boundary conditions are not symmetrical, this particular Casimir system does not exhibit temperature inversion symmetry \cite{Ravndal&Tollefsen89}, see, however, \cite{Santos&Tort1999} for a way of circunventing this limitation. 
\section{Transformations for the repulsive Casimir force}
In order to establish relations similar to (\ref{dualCasimir}) for the repulsive electromagnetic Casimir effect it is convenient to follow closely Ref. \cite{Fukushima&Ohta2001}.

Consider a pair of parallel infinite plates one of which is perfectly conducting ($\epsilon\to\infty$) and placed at $z=0$, while the other is infinitely permeable ($\mu\to\infty$) and placed at $z=\ell$. This setup was considered for the first time by Boyer \cite{Boyer1974} who showed in the framework of random electrodynamics that the distorted vacuum energy is positive and that results in a repulsive force per unit area between the plates. The simplicity of this geometrical setup allow us to simulate the electromagnetic problem by a scalar massless field. The boundary conditions then read: $\phi(t,x,y,z=0)$ and $\partial_z\phi (t,x,y,z=\ell)=0$. The eigenvalues labeling the normal modes of the field in the ${\cal OZ}$ direction are
\begin{equation}
k_z=\frac{\pi}{\ell}\left(n+\frac{1}{2}\right), 
\end{equation}
where $n \in N-1=\lbrace 0, 1, 2, ...\rbrace$. The normal modes  in the ${\cal OX}$ and ${\cal OY}$ directions are continous, i.e.: $k_x, k_y \in \cal{R}$. The unregularized vacuum energy is given by
\begin{equation}\label{vac energy}
E=\hbar c L^2\int \frac{dk_xdk_y}{\left(2\pi\right)^2}\sum_{n\in N-1}K_n,
\end{equation}
where
\begin{equation}
K_n=\sqrt{k_\bot^2+\frac{\pi^2}{\ell^2}\left(n+\frac{1}{2}\right)^2},
\end{equation}
with $k_\bot:=\sqrt{k_x^2+k_y^2}$. The discrete sum in (\ref{vac energy}) can be replaced by the integral representation
\begin{equation}
I(k_x,k_y)=-\oint\frac{dq}{2\pi}\sum_{n\in N-1}\frac{2q^2}{q^2+K_n^2},
\end{equation}
as can be easily verified with the help of the residue theorem. The vacuum energy can be written
\begin{equation}
E=\hbar c L^2\int \frac{dk_xdk_y}{\left(2\pi\right)^2}I(k_x,k_y).
\end{equation}
Now we define $z=\ell\sqrt{q^2+K_n^2}$ and make use of \cite{Gradshteyn&Ryzhik94}
\begin{equation}
\frac{\tanh z}{z}=\sum_{n\in N-1}\frac{2}{z^2+\frac{\pi^2}{\ell^2}\left(n+\frac{1}{2}\right)^2},
\end{equation}
therefore
and write
\begin{eqnarray}\label{ITANH}
I(k_x,k_y)& = & -\oint\frac{dq}{2\pi}\ell q^2\frac{\tanh{\left(\ell\sqrt{q^2+k_\bot^2}\right)}}{\sqrt{q^2+k_\bot^2}} \nonumber \\ 
& = -&\oint\frac{dq}{2\pi}\frac{\ell q^2}{\sqrt{q^2+k_\bot^2}}
+\oint\frac{dq}{2\pi}\frac{2\ell q^2}{\sqrt{q^2+k_\bot^2}\left(e^{2\ell\sqrt{q^2+k_\bot^2}}+1 \right)}.
\end{eqnarray}
The first term and the second terms in (\ref{ITANH}) can be treated exactly as in \cite{Fukushima&Ohta2001} and after subtracting the continous part we end up with a Casimir energy density given by
\begin{eqnarray}
u &=&\frac{1}{\ell}\int \frac{d^3k}{\left(2\pi\right)^3}ln\left(1+e^{-2\ell q}\right)\\ \nonumber
&=&\frac{1}{2\ell\pi^2}\int_0^\infty dq\,q^2\,\ln\left(1+e^{-2\ell q}\right)\\ \nonumber
&=& \frac{7}{8}\frac{\pi^2}{720\ell^4},
\end{eqnarray}
The Casimir pressure is
\begin{eqnarray}
p=-\frac{\partial\left(\ell u\right)}{\partial\ell}&=& 2\int \frac{d^3q}{\left(2\pi\right)^3}\frac{q}{e^{2\ell q}+1}\\ \nonumber
&=&\frac{1}{\pi^2}\int_0^\infty dq \frac{q^3}{e^{2\ell q}+1}\\ \nonumber
&=& \frac{7}{8}\frac{\pi^2}{240\ell^4},
\end{eqnarray}
where we have used the result \cite{Gradshteyn&Ryzhik94}
\[
\int_0^\infty\frac{dx\,x^{\nu-1}}{e^{\mu x}+1}=\frac{1}{\mu^\nu}\left(1-2^{1-\nu}\right)\Gamma\left(\nu\right)\zeta\left(\nu\right), 
\]
with $\Re\,\mu >0$ and $\Re\,\nu > 1$. 
The relevant (for our purposes) thermodynamical functions now are the ones derived from the fermionic partition function for a neutrino gas in thermal equilibrium with a heat reservoir at temperature $T$ and chemical potential equal to zero
\begin{equation}
p=-f=-2T\int \frac{d^3q}{\left(2\pi\right)^3}\ln\left(1+e^{-q/T}\right),
\end{equation}
and
\begin{equation}
u=-\frac{\partial p/T}{\partial \left(1/T\right)}=2\int \frac{d^3q}{\left(2\pi\right)^3}\frac{q}{e^{q/T}+1}.
\end{equation}
Where $f$ is the free energy density and is the thermal energy density. Notice that apart from an all-important algebraic sign the ratio between the Casimir pressure for Boyer's setup and the Casimir pressure for the conducting plates setup is the same as the ratio between the pressure for a massless neutrino gas and a photon gas, namely, $7/8$. It is easily seen that in this example the transformations similar to (\ref{dualCasimir}) are
\begin{equation}\label{dualBoyer}
2\ell\rightleftharpoons \frac{1}{T}, \hskip 2cm p\to u\to -p.
\end{equation}

The transformations (\ref{dualBoyer}) above also lead to an effective temperature that depends on the geometrical features of this Casimir system and does not obey the fundamental relation given by (\ref{trueT}).
\section{Conclusions}
In this brief paper we argued that a distinction between temperature inversion symmetry and the so-called duality transformations must be made. In particular we have shown in an example that it is possible to find transformations between spatial extension and temperature and between energy density and pressure in a system that does not exhibit temperature inversion symmetry. It seems that a deeper understanding of the temperature inversion symmetry depends on knowledge of the structure of the euclidean manifold and its symmetry group $O(4)$, even though as mentioned in \cite{Fukushima&Ohta2001} topological structures in euclidean space cannot always be construed as physical objects in minkowskian spacetime. It must be strongly emphasized also that the temperature inversion symmetry does more than connect the zero and the very high temperature limits. In particular, since (\ref{TIS1}) holds for arbitrary temperature, from a low temperature approximation to the Casimir effect at finite temperature we can write the high temperature one, and thus we can evaluate corrections to the Stefan-Boltzmann term that depend on the boundary conditions imposed  on the electromagnetic field.

\end{document}